\documentclass[11pt]{JHEP3}
\usepackage{latexsym}

\renewcommand{\nc}{\newcommand}
\newcommand{\rnc}{\renewcommand}

\nc{\be}{\begin{equation}}
\nc{\ee}{\end{equation}}
\nc{\bea}{\begin{eqnarray}}
\nc{\eea}{\end{eqnarray}}

\nc{\trac}[2]{{\textstyle\frac{#1}{#2}}}

\nc{\ex}[1]{\mbox{e}^{\,\textstyle#1}}

\nc{\CC}{\Bbb{C}}
\nc{\HH}{\Bbb{H}}
\nc{\PP}{\Bbb{P}}
\nc{\RR}{\Bbb{R}}
\nc{\ZZ}{\Bbb{Z}}
\nc{\II}{\Bbb{I}}
\nc{\EE}{\Bbb{E}}

\rnc{\a}{\alpha}
\rnc{\b}{\beta}
\rnc{\d}{\delta}
\nc{\ga}{\gamma}
\nc{\la}{\lambda}
\nc{\f}{\phi}
\nc{\p}{\psi}
\nc{\e}{\eta}
\rnc{\c}{\chi}
\nc{\eps}{\epsilon}
\nc{\om}{\omega}
\nc{\Om}{\Omega}

\nc{\ad}{\mathop{\mbox{ad}}\nolimits}
\nc{\tr}{\mathop{\mbox{tr}}\nolimits}
\nc{\Tr}{\mathop{\mbox{Tr}}\nolimits}
\nc{\Det}{\mathop{\mbox{Det}}\nolimits}
\rnc{\det}{\mathop{\mbox{det}}\nolimits}
\nc{\rk}{\mathop{\mbox{rk}}\nolimits}
\nc{\del}{\partial}
\nc{\diag}{\mathop{\mbox{diag}}\nolimits}
\nc{\ra}{\rightarrow}
\nc{\Ra}{\Rightarrow}
\nc{\LRa}{\Leftrightarrow}
\nc{\lra}{\leftrightarrow}
\nc{\ot}{\otimes}
\rnc{\ss}{\subset}
\nc{\nul}{\noindent\underline}
\nc{\non}{\nonumber\\}
\nc{\mat}[4]{\left(\begin{array}{cc}#1&#2\\#3&#4\end{array}\right)}
\rnc{\lg}{\frak{g}}
\nc{\G}[3]{\Gamma^{#1}_{\;{#2}{#3}}}
\nc{\nam}{\nabla_{\mu}}
\nc{\nan}{\nabla_{\nu}}
\nc{\dx}{\dot{x}}
\nc{\dxl}{\dot{x}^{\la}}
\nc{\dxm}{\dot{x}^{\mu}}
\nc{\dxn}{\dot{x}^{\nu}}
\nc{\ddx}{\ddot{x}}
\nc{\ddxm}{\ddot{x}^{\mu}}
\nc{\ddxn}{\ddot{x}^{\nu}}
\nc{\dxi}{\dot{\xi}}
\nc{\ddxi}{\ddot{\xi}}

\def\la{\label}
\def \p {\phi}

\title{On the Hagedorn Behaviour of Singular Scale-Invariant
Plane Waves}
\author{Matthias Blau\\
Institut de Physique, Universit\'e de Neuch\^atel\\
Rue Breguet 1, CH-2000 Neuch\^atel, Switzerland\\
\email{matthias.blau@unine.ch}}

\author{Monica Borunda\\
Institut de Physique, Universit\'e de Neuch\^atel\\
Rue Breguet 1, CH-2000 Neuch\^atel, Switzerland\\
\email{monica.borunda@unine.ch}}

\author{Martin O'Loughlin\\
S.I.S.S.A. Scuola Internazionale Superiore di Studi
Avanzati\\
Via Beirut 4, I-34014 Trieste, Italy\\
\email{loughlin@sissa.it}}

\abstract{As a step towards understanding the properties of string theory
in time-dependent and singular spacetimes, we study the divergence of 
density operators for string ensembles in singular scale-invariant
plane waves, i.e.\ those plane waves that arise as the Penrose limits of
generic power-law spacetime singularities. We show that the 
scale invariance implies 
that the Hagedorn behaviour of bosonic and supersymmetric strings in these
backgrounds, even with the inclusion of RR or NS fields, is the same as 
that of strings in flat space. This is in marked contrast to the behaviour 
of strings in the BFHP plane wave which exhibit quantitatively and 
qualitatively different thermodynamic properties.}

\begin{document}

\section{Introduction}

The study of string theory in the cosmological setting of non-trivial
time-dependent and possibly singular backgrounds is an outstanding
problem. While string theory in general time-dependent backgrounds is hard, 
in principle string theory in plane waves (or, say, flat orbifolds)
is exactly
solvable. Hence one can hope to gain some insight into the nature and
properties of string theory in general time-dependent backgrounds  by
studying it in the context of time-dependent plane waves (or time-dependent
orbifolds). With this motivation, in \cite{mm,mmga,bbop1,bbop2} we set out
to systematically identify tractable yet physically interesting
time-dependent plane wave metrics and develop tools for the quantisation of
string theory in such backgrounds.\footnote{For a detailed review of recent
work on time-dependent orbifolds see \cite{lcmc}.}

In \cite{bbop1,bbop2} it was found that the plane
wave metric
\be
ds^2 = 2dudv + A_{ab}(u)x^a x^b du^2 + d\vec{x}^2
\ee
associated to every space-time metric and choice of null 
geodesic in that space-time by means of the Penrose limit 
construction \cite{Penrose,bfhp2,bfp}, encodes covariant information
about the rate of growth of curvature and geodesic deviation along the
null geodesic. In particular, singularities of $A_{ab}(u)$ 
result from curvature singularities of the original space-time.

Moreover, in \cite{bbop1,bbop2} we studied the Penrose limits
of metrics with space-time singularities.
We found that the resulting plane wave metrics for a large class of 
black hole, cosmological and null singularities (namely all metrics with 
singularities of power-law type \cite{SI,CS,GP} subject to some energy
condition) exhibit a remarkably universal leading $u^{-2}$-behaviour,
\be
A_{ab}(u) \sim u^{-2}\;\;,
\label{u2}
\ee
near the singularity. Plane waves with such a wave profile are scale
invariant, i.e.\ invariant under the scaling (boost) $(u,v)\ra(\lambda
u,\lambda^{-1}v)$ and hence also homogeneous \cite{bfp,prt,mm}.
This scale invariance will turn out to play an important
role in the following.

Thus the considerations of \cite{bbop1,bbop2} single out
the singular scale-invariant plane waves with profile $\sim
u^{-2}$ as the backgrounds to consider in order to obtain insight into
the properties of string theory near physically reasonable space-time
singularities.
String theory in these singular homogeneous plane wave backgrounds
is exactly solvable \cite{deVega:1990ke,prt,mmga}, and 
various aspects of string theory in this class of backgrounds have already
been studied in particular in \cite{prt}.

Recently, motivated by the BMN correspondence \cite{bmn}, 
thermodynamical aspects of string theory in the BFHP plane wave 
\cite{bfhp1,bfhp2,rrm,mt} and other plane waves with constant 
(i.e.\ $u$-independent) profile $A_{ab}$, have been investigated 
in e.g.\ \cite{PandoZayas:2002hh}-\cite{bigazzi}.
In particular, it was found 
that the Hagedorn behaviour of strings in such backgrounds 
(and with RR flux) differs 
quantitatively and qualitatively from that of strings in flat space.

Here we wish to study the analogous question for strings in the singular
scale-invariant plane wave backgrounds with profile (\ref{u2}). In order
to address this issue we need to first come to terms with the fact that
string theory in these time-dependent singular homogeneous plane waves
leads to a time-dependent light-cone Hamiltonian. Thus the study of the
``thermodynamics'' of such systems requires some care. 

Following \cite{Kim:2000xb,Kim:2001pg,Dodonov},  
one can study the evolution of non-equilibrium systems with respect to 
invariants of the system. Given such an invariant $I$, i.e.\ an operator
which satisfies $dI/dt=0$, one can introduce an analogue 
of the Boltzmann thermal state, the density operator 
\be
\hat\rho_I=\frac{\ex{-\beta I}}{\tr \ex{-\beta I}}\;\;.
\ee
This density operator satisfies the quantum counterpart of the classical
Liouville theorem for the phase space density, namely $\tr\hat\rho_I=1$
and $d\hat\rho_I/dt=0$. 
A convenient choice of invariant for a system with a time-dependent
Hamiltonian $H(t)$ is $I=H(t_0)$, as 
it reduces to the standard
choice in the case of a time-independent system and provides an ``adiabatic''
approximation to the system provided that $H(t)$ varies sufficiently
slowly with time near $t=t_0$.

We will study the corresponding ``thermal'' partition function
\be 
Z_I(\beta) = \tr \ex{-\beta I}
\ee
for strings in the singular scale-invariant plane waves. Our main result 
is that the partition fuction for bosonic or type II strings, even with 
the inclusion of RR fields, diverges at a critical value $\beta_c$ of the
parameter $\beta$ which is identical to the inverse Hagedorn temperature
$\beta_c=1/T_H$ of strings in flat space \cite{alvarez}. This is in marked
contrast to the thermodynamical behaviour of strings in the non-singular
constant $A_{ab}$ plane waves which, as mentioned above, can be quite
different. 

The calculation
establishing this result highlights the significance and implications
of the scale invariance of these plane waves. Indeed scale invariance 
can be seen to imply 
that the string mode frequencies are independent of the light-cone momentum
and hence uniformly approach the flat space frequencies at large string mode
number $n$.

This may be an indication that string propagation in these scale invariant 
plane waves, which arise as Penrose limits of space-time singularities, 
has properties more in common with the propagation of strings in flat space
than either have with strings in the BFHP and other time-independent
plane waves.

In section 2 we summarise the approach of
\cite{Kim:2000xb,Kim:2001pg,Dodonov},  
to non-equilibrium thermodynamics via invariants.
In section 3, following \cite{prt} we present
the light-cone Hamiltonian of type IIB superstring theory in 
singular homogeneous plane wave backgrounds and exhibit its relation
to our preferred invariant. The implications of the scale invariance
are explored in section 4 and the detailed computation of the 
partition function is given in section 5.

\section{Invariants and Thermodynamics of Time-Dependent Systems}

For present purposes we will adopt the point of view (see e.g.\ 
\cite{Kim:2000xb,Kim:2001pg,Dodonov}) that suitable analogues of the Boltzmann 
thermal state for a time-independent Hamiltonian system,
\be
\hat{\rho}_H = \frac{\ex{-\beta H}}{\tr \ex{-\beta H} }\;\;,
\ee
can be constructed as density operators of the form
\be
\hat{\rho}_I = \frac{\ex{-\beta I}}{\tr \ex{-\beta I}}\;\;,
\label{rhoI}
\ee
where $I$ is an {\em invariant} of the system, i.e.\ a possibly explicitly
time-dependent operator in the Heisenberg picture satisfying the equation
\be
\frac{d}{dt}I = \frac{\partial}{\partial t}I
 + i[H,I] = 0\;\;.
\label{dti}
\ee
Note that for a time-independent system evidently the Hamiltonian $H$ itself
satisfies this equation.
In particular, the density operator satisfies the quantum 
counterpart of the classical Liouville theorem for the phase space density, 
namely $\tr \hat{\rho}_I=1$ and the Liouville-von Neumann equation
\be
\frac{d}{dt}\hat{\rho}_I = \frac{\partial}{\partial t}\hat{\rho}_I
 + i[H,\hat{\rho}_I] = 0\;\;.
\ee
As a consequence of this equation, the density operator $\hat{\rho}_I (t)$ 
allows one to calculate the time evolution
of the expectation value of 
any operator ${\cal O}$ in the mixed state characterised by the invariant
$I$,
\be
<{\cal O} >_I(t) = \tr( {\cal O} \hat\rho_I )\;\;.
\ee
Different choices of $I$ 
correspond to different initial ``thermal'' ensembles with partition
functions 
\be
Z_I(\beta) = \tr \ex{-\beta I}\;\;.
\label{ZI}
\ee
In particular, in \cite{Kim:2000xb} it has been shown that 
this reproduces and unifies various approaches to non-equlibirum 
thermodynamics such as mean-field or Hartree-Fock methods.

The parameter $\beta$ is in many ways analogous to
an inverse temperature $1/T$. However, 
here and in the following we occasionally put the word ``thermal'' in quotes
to emphasise that
we are not claiming that (\ref{rhoI}) describes
``the system at temperature $\beta=1/T$''.\footnote{Such an
identification has been used in \cite{Kim:2001pg}
to define an analogue of temperature 
for non-equilibrium systems.}

For the harmonic oscillator systems under consideration (namely the 
light-cone Hamiltonians for strings on time-dependent plane waves), 
invariants are easy to come by, e.g.\ using invariant oscillators 
(Appendix A).\footnote{See \cite{mm} for a derivation of the Lewis-Riesenfeld
theory of invariants of time-dependent harmonic oscillators \cite{LR} from
the geometry of plane waves.}

A convenient, but by no means the unique acceptable, choice is the
``instantaneous'' density operator
\be
I=H(t_0)
\ee
(which will be explicitly time-dependent when written in terms of 
Heisenberg operators). The corresponding density operator reduces to
the standard choice in the case of a time-independent system, and it
provides an ``adiabatic'' approximation to the system provided that 
$H(t)$ varies sufficiently slowly with time near $t=t_0$.

\section{The Light-Cone Hamiltonian and the Invariant}

We now summarise some results from \cite{prt} regarding the light-cone
Hamiltonian for bosonic strings in a purely dilatonic singular homogeneous
plane wave background (we will briefly discuss other string theories and
backgrounds at the end of section 5).

In the light cone gauge
\be
U(\sigma,t)=2\alpha' p_vt\;\;,
\ee
($p_v=p^u$ the light-cone momentum) 
the dynamics of the transverse string coordinates $X^a(\sigma,t)$ in
the plane wave metric
\be
ds^2 = 2du dv +A_{ab}(u)x^a x^b du^2 + d\vec{x}^2 \;\;,
\ee
is governed by the quadratic light-cone Hamiltonian
\be
H_{lc}(t)=-p_u= \frac{1}{8\pi\alpha^{\prime 2}p_v}
\int_0^\pi d\sigma\,(\dot X^{a2} +X^{\prime a2} - 4 \alpha'^2 p_v^2
A_{ab}(2\alpha'p_vt)X^aX^b)\;\;.
\ee
Note that precisely for the singular plane waves with profile
\be
A_{ab}(u) = - \omega_a^2 \delta_{ab} u^{-2}
\label{prof1}
\ee 
the dependence of the light-cone Hamiltonian on $p_v$ disappears (up to
an overall factor). This is a consequence of the {\em scale invariance}
\be
(u,v) \ra (\lambda u, \lambda^{-1}v)
\ee
characterising the corresponding homogeneous plane wave metric
\cite{mm}.

Thus the Fourier modes of the string are harmonic
oscillators with frequencies
(up to a standard overall factor of $1/\alpha'p_v$)
\be
\omega^a_n(t) = \sqrt{n^2 + \frac{\omega_a^2}{4t^2}}\;\;.
\label{omega}
\ee
We will consider the case that all the frequencies $\omega_a$ are 
real. This excludes vacuum plane waves, but includes e.g.\  
the Penrose limits of FRW metrics \cite{bfp,bbop1}
and a large class of other power-law singularities 
\cite{bbop2}.\footnote{For 
a dicussion of some aspects of ``imaginary''
frequencies in plane waves see e.g.\ \cite{bgs,dmlz}.}

The natural mode expansion in terms of solutions
to the classical equations of motion and the corresponding 
invariant oscillators $\alpha^a_n$ 
(Appendix A) leads to a non-diagonal
light-cone Hamiltonian operator, namely (cf.\ \cite{prt} and (\ref{ndh}))
\be
H_{lc}(t) = \frac{1}{\alpha'p_v}[
H_0(t) + \frac 12\sum_{n=1}^{\infty}\sum_{a=1}^d 
[\Omega_n^a(t)
(\alpha^a_{-n}\alpha_n^a +\tilde\alpha_{-n}^a\tilde\alpha_{n}^a ) -B_n^a(t )
\alpha_n^a\tilde\alpha_{n}^a-B^{a\star}_n(t
)\alpha_{-n}^a\tilde\alpha_{-n}^a]]
\label{hlc1}
\ee
where $H_0(t)$ is the Hamiltonian of the zero modes and the time-dependent
coefficients are quadratic expressions in the Bessel function solutions to
the string mode equations.

As shown in this specific example in \cite{prt}, and discussed
more generally in Appendix A in a quantum mechanical context, the
light-cone Hamiltonian operator simplifies significantly when written
in an explicitly time-dependent but diagonalising basis of oscillators
$a^a_n(t), \tilde{a}{}^a_n(t)$, 
in which the coefficients are just
the appropriately normalised classical frequencies 
(\ref{omega}) rather than the complicated functions $\Omega_n^a(t)$
appearing in (\ref{hlc1}),
\be
H_{lc}(t) = \frac{1}{\alpha'p_v}[H_0(t) + \sum_{n=1}^{\infty}\sum_{a=1}^d 
\frac{\omega^a_n(t)}{n}
(a^{a\dagger}_n(t) a^a_n(t) 
+\tilde{a}{}^{a\dagger}_n(t) \tilde{a}{}^a_n(t))] + h(t)\;\;.
\ee
Here $h(t)$ is a normal ordering c-function,
\be
h(t) = 
\sum_{a=1}^d  (\sum_{n=1}^\infty\sqrt{n^2 + \frac{\omega_a{}^2}{4t^2}} + 
\frac{\omega_a}{4t}) \;\;,
\ee
which we will discuss in more detail in section 5.
The oscillators are normalised (as are the $\alpha$ oscillators in
(\ref{hlc1})) to have the non-zero commutation relations
\bea
[a^{a}_n(t),a^{b\dagger}_m(t)] &=& n\delta_{n+m}\delta^{ab}\cr
[\tilde{a}{}^{a}_n(t),\tilde{a}{}^{b\dagger}_m(t)] 
&=& n\delta_{n+m}\delta^{ab}\;\;.
\eea

In particular, in this basis the invariant 
\be
I = H_{lc}(t_0)\;\;
\label{ihlc}
\ee
takes a very simple form and its corresponding ``thermal'' partition function
is easy to calculate (see section 5). 

\section{Implications of Scale Invariance}

As we have seen, the string mode frequencies in the scale invariant plane
wave are the $p_v$-independent but $t$-dependent
\be
\omega^a_n(t) = \sqrt{n^2 + \frac{\omega_a^2}{4t^2}}\;\;.
\ee
This highlights the special feature of the scale invariant plane waves, 
namely that for fixed $t$ their large $n$ behaviour is exactly that of 
flat space,
\be
\omega_n^a \stackrel{n\ra\infty}{\longrightarrow} n \;\;.
\ee
The above behaviour of strings in scale invariant homogeneous plane waves
should be contrasted with the behaviour of the string modes 
in the plane waves with constant wave profile
\be
A_{ab}(u) = - \mu_a^2 \d_{ab}\;\;.
\label{prof2}
\ee
In this case one finds (up to the same overall factor as in (\ref{omega})) 
the now $t$-independent but $p_v$-dependent frequencies \cite{rrm,mt}
\be
\tilde{\omega}^a_{n}(p_v) = \sqrt{n^2 + \alpha'^2p_v^2 \mu_a^2}\;\;.
\label{omegat}
\ee
While these also behave as $\sim n$ for large $n$ and fixed $p_v$, 
the integration over the light-cone momentum that arises in the calculation
of the partition function implies that for the
symmetric (i.e.\ constant $A_{ab}$) 
plane waves, or any $u$-dependence of the frequencies other
than $u^{-2}$ times a bounded function of $u$, there can be significant
departures from the flat space behaviour even for $n\ra\infty$.

This seems to indicate that the scale invariant singular homogeneous
plane waves have properties more in common with strings in flat space
than either have with the symmetric plane waves. Another manifestation
of this is the fact that for $u>0$ the $\omega_a\ra 0$ limit of the
metric with profile (\ref{prof1}) is the flat metric, while rescaling
$\mu_a$ in (\ref{prof2}) is an isometry of the metric (so that the flat
space emerging at $\mu_a=0$ is a non-Hausdorff limit).

In particular, this has direct implications for the issue we want
to study in this paper, namely the Hagedorn behaviour of strings
in scale invariant plane waves, which depends on the properties of
the string modes and the exponential growth of the number of states
at large $n$.  On the basis of the above reasoning, one might expect
this behaviour to be identical to that in flat space, and we will
confirm this by an explicit calculation in the next section. On the
other hand, as was already mentioned in the Introduction, the
symmetric plane waves exhibit a  different behaviour.

\section{The ``Thermal'' Partition Function}

In this section we will present the calculation of the thermal partition
function (\ref{ZI})
corresponding to the invariant $I=H_{lc}(t_0)$ (\ref{ihlc}). 
We will add to the invariant the light-cone momentum $p_v$ in such 
a way that in the flat space limit the density operator 
reduces to the standard expression for 
the time-like Hamiltonian $(p_u + p_v)/\sqrt{2}$.

Thus the density operator of interest is 
\be
\hat{\rho}_I = \frac{\ex{-\frac{\beta}{\sqrt{2}} (I+ \hat{p}_v)}}
{\tr \ex{-\frac{\beta}{\sqrt{2}} (I+ \hat{p}_v)}}\;\;,
\ee
and we will now study the associated thermal partition function
\be
Z_I(\beta)=\tr \ex{-\frac{\beta}{\sqrt{2}} (\hat{p}_v+ I)}\;\;.
\ee
To evaluate this trace, we need to implement the level matching condition.
We first do the calculation for a plane wave
solution with non-trivial dilaton and no other non-trivial fields
in the bosonic string theory  \cite{prt} and will discuss later the
generalisation to other string theories and field configurations. 

In terms of either of the oscillator bases the level 
matching condition is simply the difference between the left and
right number operators, 
\be
N - \tilde{N} = \sum_{a=1}^d
\sum_{n=1}^\infty (\alpha_{-n}^a\alpha_n^a - \tilde{\alpha}_{-n}^a
\tilde{\alpha}_n^a) = \sum_{a=1}^d
\sum_{n=1}^\infty (a_n^{a\dagger} a_n^a - 
\tilde{a}{}_n^{a\dagger}
\tilde{a}{}_n^a)\;\;.
\ee

Thus $Z_I(\beta)$ is
\be
Z_I(\beta )=\int dp_v \int d\lambda\; \ex{-\frac{\beta p_v}{\sqrt{2}}} \, 
\tr \ex{-\frac{\beta}{\sqrt{2}} I+2\pi i\lambda (N-\tilde{N})}\;\;.
\ee
For notational simplicity we will from now on consider the case where all the
frequencies are equal,
$\omega_a=\omega$, but our final result for the critical 
(Hagedorn) value of $\beta$ 
turns out to be independent of $\omega$ and is hence, in particular, also
valid when the frequencies are distinct. 
We will also abbreviate $\omega_n(t_0)\equiv \omega_n$. 

In terms of the complex variable $\tau$,
\be
\tau=\tau_1 +i\tau_2 =\lambda +i\frac{\beta }{2\sqrt{2}\alpha'\pi p_v}\;\;,
\ee
the trace inside the integral is
\bea
\tr \ex{-2\pi\tau_2 I+2\pi i\tau_1 (N\!-\!\tilde{N})} &\!=\!&
\prod_{n=1}^\infty
|\sum_{m=0}^\infty \ex{(-2\pi\tau_2 \omega_n + 2\pi i\tau_1 n)m}|^{2d} \non
&& \times
\sum_{m=0}^\infty \ex{-2\pi\tau_2\omega_0 dm} \ex{-2\pi\tau_2 h(t_0)}\non
&=& (\prod_{n=-\infty}^{\infty} 
\frac{1}{1- \ex{-2\pi\tau_2 \omega_n + 2\pi i\tau_1 n}})^d
\ex{-2\pi\tau_2 h(t_0)}\non
&=&
D_{0,0}(\tau_1,\tau_2;\frac{\omega}{2t_{0}})^{-d}\;
\ex{2\pi\tau_2(d\Delta_0(q)-h(t_0))}
\label{trace}
\eea
where $D_{0,0}(\tau;q)$ is a generalised (massive) theta function 
(see Appendix B for the definition).

The normal ordering term 
\be
h(t_0)= d  (\sum_{n=1}^\infty\sqrt{n^2 + \frac{\omega^2}
{4t_0^2}} + \frac{\omega}{4t_0})\;\;,
\label{hto}
\ee
is divergent. Actually $h(t_0)$ has 
two divergences, the first one arising from a sum over $n$ of $n$  - a 
quadratic divergence, the second one arising from a sum over $n$ of $1/n$ - 
a logarithmic divergence. In the present case (generalising 
bosonic string theory in Minkowski background) this quadratic 
divergence is cancelled by 
a generalised zeta function regularisation.  The details of the resummation 
are given in Appendix B with the result that ($q=\omega/2t_0$),
\bea
\frac{h(t_0)}{d} &=& \frac{1}{2}\sum_{n\in Z}\sqrt{n^2 + q^2}\\
&=& -\frac{q}{\pi} \sum_{k=1}^\infty \frac{K_1(2\pi kq)}{k}
- \frac{q^2}{4} \Gamma(-1)\\
&\equiv& \Delta_0(q) - \frac{q^2}{4}\Gamma(-1)\;\;,
\eea
where $K_1$ is a modified Bessel function of the second kind.

For $q=0$ (Minkowski space) it is easy to see that we get the
usual result of zeta function regularisation that $h(t_0) = -d/12$ (see
Appendix B). However, for $q \neq 0$
the term proportional to $q^2$ needs interpreting as it is infinite. 
The source of this infinity is 
the subleading logarithmic divergence $\sim \zeta(1)$ of 
$h(t_0)$ mentioned above, as can also be seen directly by
expanding and resumming the series for small $q$  by 
utilising the binomial expansion of the square root (\ref{zetaf}).

This remaining divergence is  more subtle, but before we consider it
in more detail we first make the following observation. Since, in the
present context, $q$ is not related to the modular parameter, i.e.\
independent of $p_v$ because of scale invariance, the relevant modular
transformation rule of $D_{0,0}(\tau;q)$ 
is \footnote{This is different from the modular
transformation for the BFHP
plane wave, for which $q$ depends on $p_v$ (\ref{omegat}) and
one has $D_{0,0}(\tau;q) = D_{0,0}(-1/\tau;q/|\tau|)$ instead, 
cf.\ the discussion in \cite{PandoZayas:2002hh}.}
\be
D_{0,0}(\tau_1,\tau_2;q) = D_{0,0}(-\tau_1/|\tau|^2, \tau_2/|\tau|^2; q|\tau|)
\;\;.
\label{dmod}
\ee
As a consequence $\tau_2 q^2$ is itself modular invariant, 
\be
\tau_2 q^2 = (\tau_2/|\tau|^2) (q|\tau|)^2\;\;,
\ee
and thus any term in $h(t_0)$ that is proportional to $q^2$ makes a
contribution to (\ref{trace}) that is modular invariant all by itself.

In principle there can 
be a logarithmic divergence arising in a two-dimensional
non-linear sigma model 
though it generally vanishes on shell or can be removed by a singular 
field redefinition as such models are formally (1-loop) scale invariant. 
In string theory one requires 
that the quantisation preserves modular invariance
and divergences are again regularised
via singular field redefinitions and appropriate subtractions (for a
discussion of these issues in the present context, see \cite{prt}). 
For us at the moment it is important simply to note that any subtraction 
that removes the divergence will be proportional to $q^2$ and thus 
will not destroy the modular covariance of the state sum.

Putting everything together, and setting $d=24$ for the bosonic string,  
we see that we can now write $Z_I(\beta)$ as
\be
Z_I(\beta)
=\int_0^\infty \frac{d\tau_2}{\tau_2^2}
\int_{-1/2}^{1/2}d\tau_1\,
\ex{-\frac{\beta^2}{4\pi\alpha^\prime \tau_2}} \,D_{0,0}(\tau_1,\tau_2;
\frac{\omega}{2t_0})^{-24}\, .
\label{zib}
\ee
The potential 
divergence in this integral arises from the region where $p_v\to\infty$, 
corresponding to $\tau_2\to 0$, and we can set $\tau_1 =0$ to determine
the leading behaviour of the integral. 

To easily determine the behaviour of the integrand one can use 
the modular transformation property (\ref{dmod}) of $D_{0,0}$ to deduce
that
for $\tau_1=0$ and in the limit $\tau_2\to 0$ the transverse partition function
behaves as
\be
D_{0,0}(\tau_1,\tau_2;\frac{\omega}{2t_0})^{-24}
\to  \exp[-\frac{2\pi}{\tau_2}\, 24\Delta_0 (\frac{\omega}{2t_0}
\tau_2)]\;\;. 
\ee
Combining this with the expansion 
of $\Delta_0$ (Appendix B), 
\be
\Delta_0(\frac{\omega}{2t_0}\tau_2)= -\frac{1}{12}+\frac 12\frac{\omega}{2t_0}
\tau_2 \;\;,
\ee
one finds that the leading behaviour of the integrand as $\tau_2 \ra 0$ is
\[ \exp[\frac{1}{4\pi \alpha^\prime\tau_2}(16\pi^2 \alpha^\prime - \beta^2)]
\;\;.\]
Thus $Z_I(\beta)$ diverges for $\beta \leq \beta_c$, with
\be
\beta_c^2=16 \pi^2\alpha^\prime. 
\ee
This is precisely the value $\beta_c=\beta_H$ 
that corresponds to the Hagedorn temperature for strings in flat space
\cite{alvarez}. 
Note that, in particular, this result is independent of $\omega$ or, 
more generally, of the frequencies $\omega_a$ determining the plane wave
background. It is also manifestly independent of $t_0$ or, in other words,
independent of the choice of invariant $I=H(t_0)$ determining the thermal
ensemble.

For the type II superstring theories, we have a similar dilatonic
background. The bosonic contribution to the ``thermal'' partition function
$Z_I(\beta)$ is as above, with $d=8$.
To include the fermionic contribution we note that in the light cone gauge
the fermionic Lagrangian is given by
\be
L_F=\psi^i\partial_{\bar z}\psi^i+\bar\psi^i\partial_z\bar\psi^i\, ,
\ee
so the fermions do not couple to the parameters of the metric, and
therefore its contribution to 
$Z_I(\beta)$ will be the same as in a flat background. It follows
that $Z_I(\beta)$ diverges for $\beta \leq \beta_c$ where
$\beta_c = \beta_H = 2\pi \sqrt{2\alpha'}$ coincides 
with the inverse Hagedorn temperature of type II strings in flat space
\cite{alvarez}.

For superstring backgrounds that also have non-vanishing form fields one 
finds the same results. For example, 
consider a type IIB superstring propagating in an homogeneous
plane wave metric supported by a Ramond-Ramond 5-form background
(the singular analogue \cite{prt,mm} of the BFHP background)
\bea
ds^2&=&2dudv -\frac{\omega^2}{u^2}\vec{x}^2 du^2+d\vec{x}^2\non
&&F_{u1234}=F_{u5678}=2\frac{\omega}{u}\;\;.
\eea
As in \cite{rrm,mt}, the GS 
light-cone gauge action for a string in this background is
\bea
L&=&L_B+L_F\nonumber\\
L_B&=&\frac 12 (\partial_+ X^a\partial_-X^a-\frac{\omega^2}{t^2}X^{a2} )\non
L_F&=&i(\theta^1\bar\gamma^-\partial_+\theta^1 
+\theta^2\bar\gamma^-\partial_-\theta^2-2\frac{\omega}{t}
\theta^1\bar\gamma^-\Pi\theta^2).
\label{lagfermions}
\eea
Notice that once again all $p_v$-dependence has disappeared. The mode
equations for the fermions can also be solved in closed form (in terms of
Bessel functions). Just as for the bosonic modes, 
their large $n$ behaviour uniformly aproaches that of
flat space. Thus the resulting critical ``inverse temperature'' $\beta_c$ is
independent of $\omega$ and equal to the type II Hagedorn temperature
of flat space,
in contrast to the result for the BFHP (and other symmetric)
plane waves \cite{PandoZayas:2002hh}-\cite{Grignani:2003cs}.
For other (e.g.\ NS) backgrounds, the string mode equations may be more
complicated (and difficult to solve in closed form), but the large $n$ behaviour
will always be as above, leading to the same conclusions regarding the
value of $\beta_c$.

\subsection*{Acknowledgements}

We thank A. Tseytlin for correspondence. 
This work has been supported by the Swiss National Science
Foundation and by the Commission of the European Communities
under contract MRTN-CT-2004-005104. MBo thanks CONACyT.

\appendix

\section{Invariant vs.\ Diagonalising Oscillator Bases}

Consider the time-dependent harmonic oscillator with Hamiltonian
\be
H = \frac{1}{2m} p^2 + \frac{m\omega(t)^2}{2} x^2\;\;.
\ee
For a time-{\em in}dependent harmonic 
oscillator one can easily find an oscillator 
basis which gives a {\it diagonal} Hamiltonian and such that the oscillators 
themselves are {\it invariant}. For a time-dependent oscillator one
cannot find a basis for which these two conditions are simultaneously 
satisfied.  

It is easy to find invariant oscillators: one defines $\alpha$ and
$\alpha^\dagger$ by the ``mode expansion''
\bea
\hat{x}(t) &=& \alpha z(t) + \alpha^\dagger z(t)^*\\
\hat{p}(t) &=& m(\alpha \dot{z}(t) + \alpha^\dagger \dot{z}(t)^*)
\label{modes1}
\eea
where $z(t)$ is a complex solution to the equations of motion 
\be
\ddot{z}(t) = -\omega(t)^2 z(t)
\ee
with the Wronskian of $z(t)$ normalised to 
\be
W(z,z^*) = z(t)\dot{z}^*(t) - z(t)^* \dot{z}(t) = \frac{i\hbar}{m}\;\;.
\ee
The Heisenberg operator equations of motion imply that indeed $\alpha$
is invariant, in the sense that it satisfies (\ref{dti}),
and the Wronskian normalisation 
condition implies that these oscillators
satisfy the standard canonical commutation relations
\be
{}[\alpha,\alpha^{\dagger}]=1\;\;.
\ee
Note that then also any, say, quadratic function of these oscillators 
with time-independent coefficients is an invariant in the sense of
(\ref{dti}).

In this basis the Hamiltonian takes the general form 
\be
\hat{H}=\frac{m}{2}[(\alpha\alpha^{\dagger} +
\alpha^\dagger\alpha)
(|\dot{z}|^2 + \omega^2 |z|^2) + \alpha^2(\dot{z}^2 + \omega^2 z^2)
+ \alpha^{\dagger 2}(\dot{z}^{*2} + \omega^2 z^{*2})]\;\;.
\label{ndh}
\ee
It is non-diagonal for $\omega(t)$ not constant, because
\be
\frac{d}{dt}(\dot{z}(t)^2 + \omega(t)^2 z(t)^2) = 2 \omega(t)\dot{\omega}(t)
z(t)^2
\ee
is then not zero, so a fortiori $\dot{z}^2 + \omega^2 z^2$ itself cannot be
zero.

For $\omega(t)=\omega$ constant, the Hamiltonian
is diagonal and explicitly the above mode expansion reads
\bea
\omega(t)=\omega \Ra &&
\hat{x}(t)=
\sqrt{\frac{\hbar}{2m\omega}}(\alpha\ex{-i\omega t} + \alpha^\dagger
\ex{i\omega t}) \non
&&\hat{p}(t)=
i\sqrt{\frac{\hbar m\omega}{2}}(-\alpha\ex{-i\omega t} + \alpha^\dagger
\ex{i\omega t})\;\;.
\label{modes}
\eea
Since the proof that the Hamiltonian in this basis is diagonal is purely
algebraic, i.e.\ does not depend on the $t$-independence of $\omega$,
we are thus led to define, in general, an alternative oscillator basis
$a$ and $a^\dagger$ by
\bea
\hat{x}(t) &=& \sqrt{\frac{\hbar}{2m\omega(t)}}
(a\ex{-i\omega(t)t} + a^\dagger \ex{i\omega(t)t})\\
\hat{p}(t) &=& i\sqrt{\frac{\hbar m\omega(t)}{2}}
(-a\ex{-i\omega(t)t} + a^\dagger \ex{i\omega(t)t})\;\;.
\eea
This indeed defines the oscilators $a$ and $a^\dagger$, as one can solve for
them in terms of
$\hat{x}(t)$ and $\hat{p}(t)$, and this determines the non-trivial
time-dependence of $a(t)$ and $a^\dagger(t)$. Nevertheless one has the
time-independent canonical commutation relations 
\be
{}[a(t),a^\dagger(t)] = 1\;\;.
\ee
In terms of these oscillators the Hamiltonian is diagonal, 
\be
\hat{H}(t) = \frac{\hbar\omega(t)}{2}(aa^\dagger 
+ a^\dagger a)
\ee
and, in particular, the frequency in the diagonal basis is just the
classical frequenncy $\omega(t)$. 

Explicitly, the time-dependent $SU(1,1)$ transformation 
( = invariance group of the oscillator algebra) from $\alpha$ to
$a$ is 
\be
a = f(t) \alpha + g(t) \alpha^\dagger\;\;,
\ee
where
\bea
f(t)&=& \sqrt{\frac{m}{2\hbar\omega(t)}}\ex{i\omega(t)t}(\omega(t)z(t) +
i \dot{z}(t))\non
g(t)&=& \sqrt{\frac{m}{2\hbar\omega(t)}}\ex{i\omega(t)t}(\omega(t)z(t)^*
+
i \dot{z}(t)^*)\;\;,
\eea
satisfy the $SU(1,1)$ condition
\be
|f(t)|^2 - |g(t)|^2 = -\frac{mi}{\hbar}W = 1
\ee
as a consequence of the condition on the Wronskian of $z(t)$.

\section{Generalised Theta Functions}

The generalised Theta function is given by \cite{takayanagi}
\be
D_{b_1,b_2}(\tau_1,\tau_2;q)\equiv \ex{2\pi\tau_2\Delta_{b_1}(q)}
\prod_{n=-\infty}^\infty
(1-\ex{2\pi\tau_2\sqrt{(n+b_1)^2+q^2}+2\pi i\tau_1(n+b_1)-2\pi i b_2})\, ,
\ee
where
\be
\Delta_b(q)\equiv -\frac{q}{\pi}\sum_{p=1}^\infty \frac{\cos (2\pi bp)}{p}K_1(2\pi qp)\, ,
\label{Delta}
\ee
and $K_1$ is a modified Bessel function of the second kind. 
Its  modular properties  are
\be
D_{b_1,b_2}(\tau_1,\tau_2;q)=
D_{b_2,-b_1}(-\frac{\tau_1}{|\tau |^2},\frac{\tau_2}{|\tau |^2};
q|\tau |)=D_{b_1,b_2+b_1}(\tau_1+1,\tau_2;q)\, .
\ee 
In the detailed calculations of this paper we only need 
$D_{0,0}(\tau_1,\tau_2;q)$ and therefore just $\Delta_0(q)$, whose
relation to the normal ordering c-function $h(t_0)$ (\ref{hto})
we will now explain.

To that end consider $F=2h(t_0)/d$, which we write as
\be
F = \sum_{n\in Z}\sqrt{n^2 + q^2}
\ee
where $q = \omega/2t_0$.
In zeta function regularisation the first step we take is to use 
Poisson resummation. Then
\be
F = \sum_{k\in Z}\int_{-\infty}^{\infty} dy\; \ex{-2\pi iky}\sqrt{y^2 + q^2}.
\ee
Rewriting the square root term as an integral, 
\be
\sqrt{y^2 + q^2} = \frac{1}{\Gamma(-1/2)} \int_0^\infty
dt\; t^{-3/2} \ex{-t (y^2 + q^2)}
\ee
we can then write
\bea
F &=& \frac{1}{\Gamma(-1/2)} \sum_{k\in Z} \int_{-\infty}^\infty
dy\; \ex{-2\pi iky} \int_0^\infty dt \;t^{-3/2}\ex{-t(y^2+q^2)}\non
&=& -\frac{2q}{\pi} \sum_{k=1}^\infty \frac{K_1(2\pi kq)}{k}
- \frac{q^2}{2} \Gamma(-1)\non
&=& 2\Delta_0(q) - \frac{q^2}{2} \Gamma(-1)\;\;.
\eea
Thus we find
\be
\frac{h(t_0)}{d} =\Delta_0(q) - \frac{q^2}{4}\Gamma(-1)\;\;.
\ee
Alternatively the binomial expansion of $F$ for small $q^2$ is
\bea
F &=& q + 2\sum_{n=1}^\infty \sqrt{n^2 + q^2}\non
&=& q + 2\sum_{n=1}^\infty \sum_{k=0}^\infty \frac{\Gamma(3/2)}{\Gamma(3/2-k)}
\frac{q^{2k}}{n^{2k-1}k!}\non
&=& q + 2\sum_{k=0}^\infty \frac{\Gamma(3/2)\zeta(2k-1)}{\Gamma(3/2-k)k!}
q^{2k}\non
&=& q + 2\zeta(-1) + q^2\zeta(1) + 2 \sum_{k=2}^\infty 
\frac{\Gamma(3/2)\zeta(2k-1)}{\Gamma(3/2-k)k!} q^{2k}\;\;,
\label{zetaf}
\eea
giving the expansion
\be
\frac{h(t_0)}{d} 
= -\frac{1}{12}+\frac q2  + \frac{q^2}{2}\zeta(1) +
\sum_{k=2}^\infty \frac{(-)^k\Gamma (k-\frac 12)}{\Gamma (-\frac 12)\Gamma 
(k+1)}\zeta(2k-1)q^{2k}
\label{Deltaexpansion}\;\;.
\ee

\rnc{\Large}{\normalsize}

\end{document}